\documentclass[conference]{IEEEtran}
\IEEEoverridecommandlockouts
\usepackage{cite}
\usepackage{amsmath,amssymb,amsfonts}
\usepackage{algorithm}
\usepackage{algorithmic}
\usepackage{graphicx}
\usepackage{textcomp}
\usepackage{xcolor}
\def\BibTeX{{\rm B\kern-.05em{\sc i\kern-.025em b}\kern-.08em
    T\kern-.1667em\lower.7ex\hbox{E}\kern-.125emX}}
\begin{document}

\title{Global Message Ordering using Distributed Kafka Clusters}

\author{

\IEEEauthorblockN{Shashank Kumar}
\IEEEauthorblockA{
\textit{University of Florida} \\
FL, USA \\
sh.kumar@ufl.edu}

\and

\IEEEauthorblockN{Aryan Jadon}
\IEEEauthorblockA{
\textit{
San Jose State University}\\
CA, USA \\
aryan.jadon@sjsu.edu
}
\and
\IEEEauthorblockN{Sachin Sharma}
\IEEEauthorblockA{
\textit{
University of Colorado Boulder}\\
CO, USA \\
sachin.sharma@colorado.edu
}
}

\maketitle
\begin{abstract}

In contemporary distributed systems, logs are produced at an astounding rate, generating terabytes of data within mere seconds. These logs, containing pivotal details like system metrics, user actions, and diverse events, are foundational to the system's consistent and accurate operations. Precise log ordering becomes indispensable to avert potential ambiguities and discordances in system functionalities. Apache Kafka, a prevalent distributed message queue, offers significant solutions to various distributed log processing challenges. However, it presents an inherent limitation: while Kafka ensures the in-order delivery of messages within a single partition to the consumer, it falls short in guaranteeing a global order for messages spanning multiple partitions. This research delves into innovative methodologies to achieve global ordering of messages within a Kafka topic, aiming to bolster the integrity and consistency of log processing in distributed systems. Our code is available on GitHub - https://github.com/aryan-jadon/Distributed-Kafka-Clusters.

\end{abstract}


\section{Introduction}

Since the inception of Web 2.0 and the ongoing evolution to Web 3.0, there has been a significant proliferation in the decentralization of computer systems\cite{joyce1987monitoring}. Concurrently, the landscape of data generation has undergone transformative shifts\cite{jadon2023overview}. Spurred by the surge of IoT(Internet of Things) devices, the prevalence of social media platforms, the rise of online services, and the myriad of digital infrastructures and architectures, there has been a meteoric surge in the magnitude, pace, and diversity of data generated\cite{10215825}. Within a nodal cluster, data emanates from multifarious sources: 

\begin{enumerate}

\item Events denoting user activities such as logins, content access, user engagements, and transactions.

\item System-centric metrics encompassing service engagements, network metrics\cite{10205730}, and node-specific resource utilizations like heap memory, CPU, and disk performance.

\end{enumerate}

In traditional systems, data analysis was predominantly conducted offline, extracting logs from operational servers\cite{Bug_Report}. In contrast, contemporary systems place significant emphasis on real-time data analysis, leveraging immediate feedback to inform subsequent operational decisions\cite{patil2023comparative}.


Apache Kafka\cite{kreps2011kafka} has evolved as a potent tool to confront the intricacies introduced by the surge in data volume. It facilitates the dependable, scalable, and proficient acquisition, preservation, and analysis of streaming data. The distributed nature of Kafka supports horizontal expansion, distributing data over numerous brokers, thus ensuring high-capacity and fault-resilient data handling\cite{ganesan2017redundancy}.

Data can be introduced into Kafka topics via diverse methods including producers, connectors, or alternative data integration techniques. Once within Kafka, this data can undergo processing, and transformation, and be accessed by an array of applications, infrastructures, or analytical workflows. Kafka's capacity to manage substantial data loads, ensure fault resilience, and facilitate real-time operations positions it as a preferred option for an extensive range of applications such as data pipeline construction, event-centric architectures, log consolidation, and stream analytics, among others\cite{nasiri2019evaluation}.

A notable challenge in utilizing Kafka is its provision of ordering guarantees limited to an individual partition and not across multiple partitions. Each Kafka partition is designated to a particular broker and functions autonomously, facilitating parallel computations and enhancing scalability\cite{le2017performance}. Yet, due to this segmented structure, it's not feasible to ensure a universal order for data spanning all partitions within a topic. This absence of comprehensive ordering across partitions poses constraints in situations necessitating rigorous sequential processing or specific event sequencing.

In this research paper, we aim to tackle the issue of attaining a universal data order across partitions within Apache Kafka using \textbf{Aggregator and Sorter}, \textbf{Single Consumer within a consumer group}, and \textbf{Batch Commit and Broadcast Protocol Algorithms}. By overcoming this limitation, our research strives to make a significant contribution to both the Kafka community and practitioners dealing with situations where maintaining a global data order is paramount for their data processing workflows. We possess assurance that our findings will not only enhance the capabilities of Kafka but also unveil novel prospects for applications that require exact event sequencing and effective dependency management across multiple partitions.

The structure of this paper is organized as follows: Section II delves into the related work, while Section III elaborates on the Proposed Architecture and Design implementations. Experimental findings are presented in Section IV, and Section V concludes the paper and offers insights into future work.

\section{Related Work}

The challenge lies in the realm of distributed systems, spanning thousands of components scattered across the globe. This complex landscape necessitates a dedicated Middleware infrastructure. These distributed systems, by their very nature, are rigid and static, requiring a transformation from point-to-point synchronous applications to large-scale asynchronous systems. This transition is pivotal due to the glaring problem: traditional setups, such as Meta Scribe, employ log aggregators that funnel data from frontend machines over sockets, eventually storing it in HDFS for offline processing\cite{karpathiotakis2019scribe}. However, this approach leaves the potential of real-time data utilization untapped, creating a substantial gap.

While other messaging queue systems, like IBM Websphere, offer global message ordering\cite{aranha2013ibm}, they falter in high throughput scenarios due to the stringent delivery guarantees mandating message exchange acknowledgments. Such guarantees, while valuable in certain contexts, prove excessive for noncritical log data. Similarly, messaging services like RabbitMQ\cite{ionescu2015analysis} and ActiveMQ\cite{snyder2011activemq} maintain global ordering but stumble when faced with the scale of data, as they lack the ability to send multiple messages within a single request, resulting in costly TCP/IP round trips.

Kafka protocol is a game-changer in the real-time data processing realm. It empowers consumers to access messages as soon as Brokers publish them\cite{wang2015building}. Kafka's pull mechanism for data access ensures consumers remain unfazed by high network traffic, thus delivering unparalleled throughput. The magic behind Kafka's success lies in its elegant architecture, leveraging Zookeeper for essential distributed tasks like data partition replication, leader consensus, and maintaining consumer and broker registries, including tracking consumer data offsets for each partition.

\begin{figure}[htbp]
\includegraphics[width=70mm,scale=0.5]{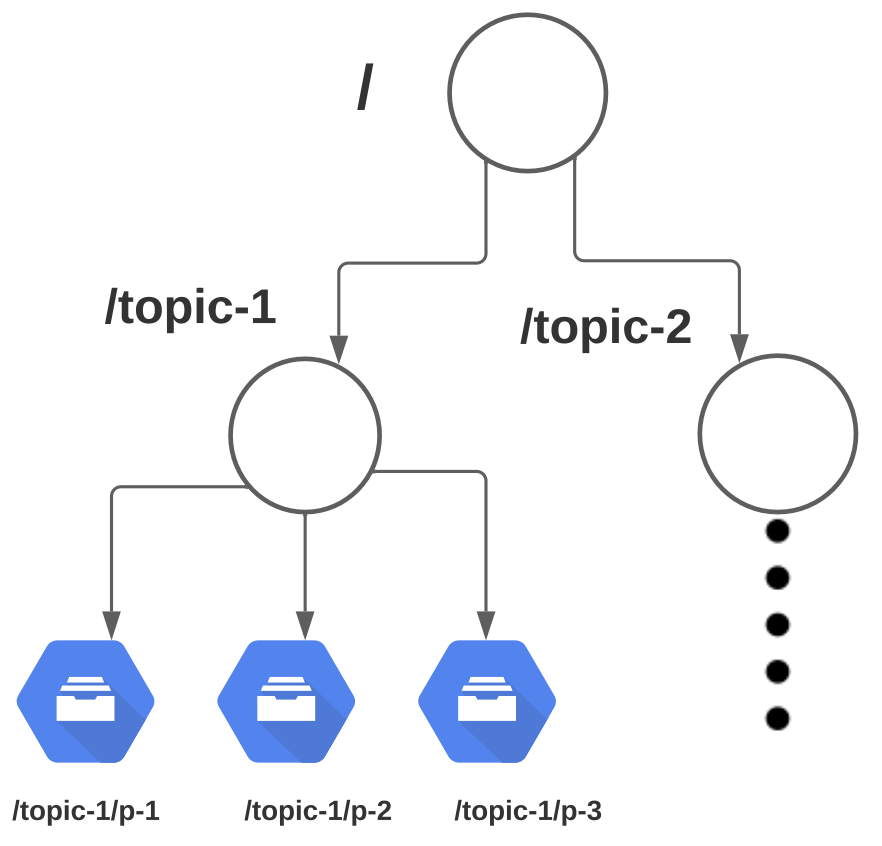}
\caption{A Standard file system for Zookeeper Namespace}
\label{fig1}
\end{figure}

Figure 1 vividly illustrates the Zookeeper's role in this architecture, organizing registries in a file directory structure. Brokers, consumers, and ownership registries are ephemeral, ensuring seamless load balancing when servers are added or removed. In addition, Kafka maintains a persistent offset registry for data recovery in case of consumer failures.

Kafka's innovative solution hinges on parallel data streaming using partitions, where messages within a partition maintain their order, enhancing throughput. However, this approach poses a challenge for applications requiring global message ordering when dealing with messages from the same topic distributed across different partitions.

LinkedIn, for instance, has deployed a Kafka library cluster within a data center to facilitate offline analysis, leveraging HDFS for delivering analytical insights\cite{jadon2022comprehensive}. Although this setup caters to applications relying on message sequencing, it primarily operates offline. In certain scenarios, data undergoes preprocessing before reaching the producer application, introducing an additional layer of complexity in the development process. Our mission is clear: harness Kafka's distributed parallel data processing and high throughput, in tandem with its streaming queue capabilities, to bridge the gap and ensure coherent message sequencing across partitions, unlocking the true potential of real-time data utilization.

\section{Proposed Architecture and Design Implementations}

To address the mentioned use cases, the creation of partitions may necessitate the use of non-intuitive keys. Furthermore, partitions limit consumption to a single consumer node. Our objective is to maintain a universal order, irrespective of the number of consumers involved.

Numerous messaging technologies, including AMQP (Advanced Message Queuing Protocol) \cite{naik2017choice} and JMS \cite{vinoski2006advanced}, offer support for Message Prioritization. These technologies enable messages to be consumed or processed in varying orders, depending on their significance.

For instance, consider a scenario in applications where customer queries need attention, and a business may need to handle the most critical cases first. Kafka, originally designed as an event streaming platform, lacks essential features such as message prioritization. To bridge this gap, we intend to introduce an intermediary layer between consumers and brokers that can facilitate message prioritization.

\subsection{\textbf{Architecture and Design}}

A concise overview of Kafka's architecture includes brokers with topics and partitions, engaging with producers and consumers. In this section, we will delve into crucial aspects of these components, laying the foundation for our upcoming architectural design.

\subsubsection*{\textbf{Producer}}

In Kafka, the producer's role involves disseminating messages among various partitions. The quantity of partitions within a topic is established during its creation. By default, the \textbf{Partitioner} utilizes a hash function of the message key to determine the appropriate partition for the message.

\subsubsection*{\textbf{Consumer}}

A consumer group subscribes to the topics it intends to receive messages from. Within each consumer group, partitions are allocated to different consumers to ensure that each partition is processed by a single consumer. The logic responsible for assigning partitions to consumers is implemented by the \textbf{Assignors}.

\subsubsection*{\textbf{Broker}}

In Kafka, each broker is referred to as a \textbf{bootstrap server}, and a Kafka cluster comprises multiple such brokers (servers). Every broker is uniquely identified by an integer ID and houses specific topic partitions. What makes Kafka intriguing is that, at any moment, a client needs to establish a connection with just one broker, and that connection provides access to the entire cluster. 

Each broker possesses knowledge of all other brokers, topics, and partitions through the maintenance of metadata across the server ensemble. The orchestration of all brokers is a key function carried out by Zookeepers. They maintain a comprehensive list of all brokers and are also responsible for orchestrating leader elections for partitions.

We are presenting three distinct designs to attain message ordering and subsequently assessing their impact on performance in comparison to the existing Kafka implementation.

\subsection{\textbf{Aggregator and Sorter Algorithm}}

In this approach, we propose a method to ensure message ordering by buffering and sorting the messages. We can use the message key field found in the ProducerHeader.java file, which assigns a unique identifier to each message, for this purpose. When multiple partitions are in use, consumers must maintain a buffer that contains messages from all partitions. If only one consumer were used, a local cache could suffice.

However, in high-load scenarios where multiple consumers are needed, each reading from a single partition, this buffer must be positioned outside the consumer layer. The middleware layer will then sequentially poll messages from the consumers and arrange them in the correct order.

For instance, if messages arrive out of order, the middleware will only deliver those that are in sequence, while retaining out-of-order messages in the buffer until the missing sequences arrive. Another approach to maintaining sequential delivery is to poll messages in sequence. Although this eliminates the need for buffering messages, it can be extremely slow.

While the Aggregator and Sorter approach effectively addresses the global message ordering issue, it compromises parallelism in a distributed system. Additionally, there is no predefined limit on buffer size. If a consumer processes messages slowly, it must either continue buffering messages or wait until the missing messages arrive. Figure 3 explains the Proposed Design using the Aggregator and Sorter Mechanism.

\begin{figure}[htbp]
\includegraphics[width=90mm,scale=0.8]{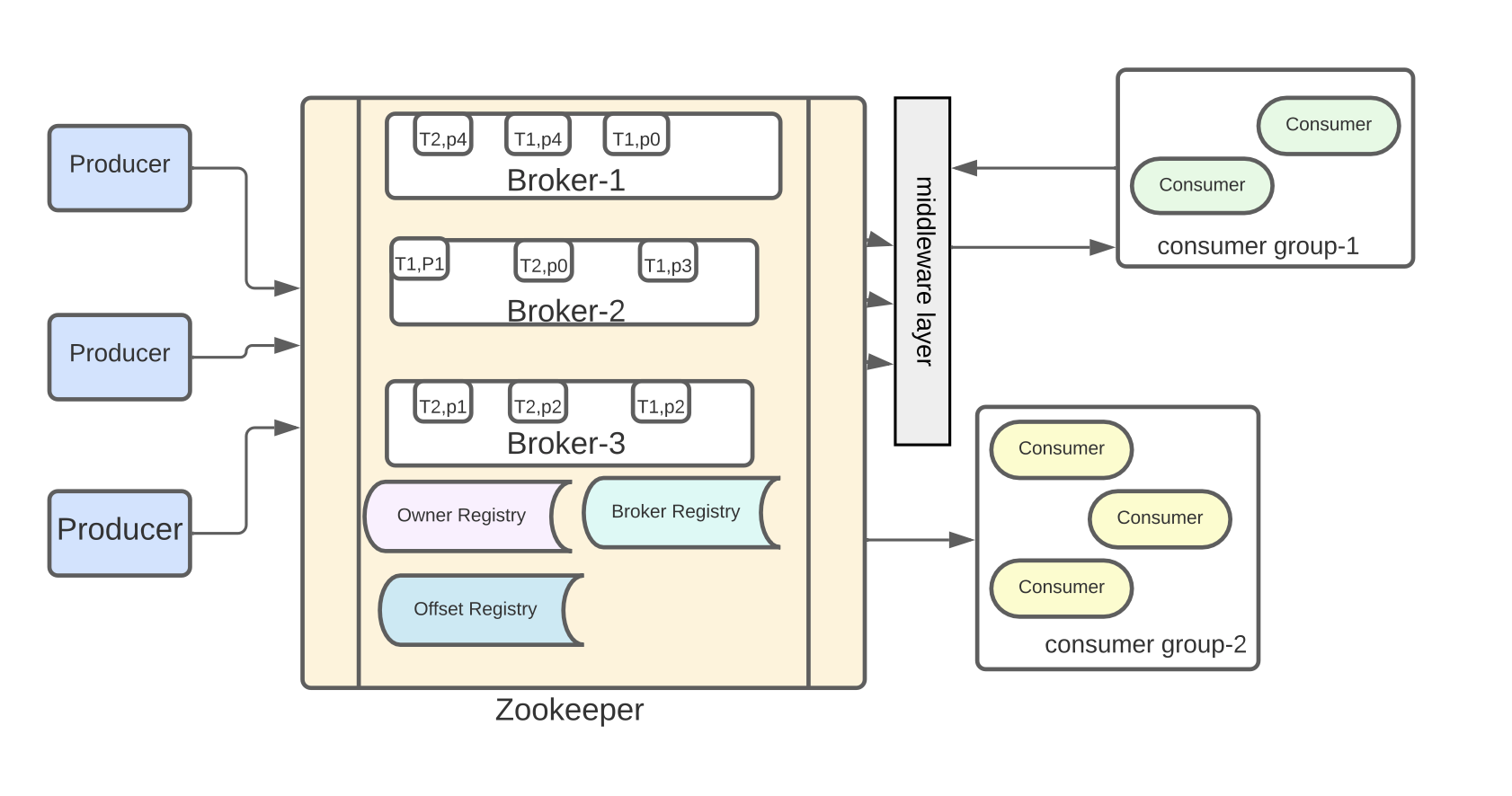}
\caption{Proposed Design using Aggregator and Sorter Mechanism}
\end{figure}

\begin{algorithm}
\label{Aggregator and Sorter}
\caption{Aggregator and Sorter}
\begin{algorithmic}
    \STATE Input: Consumers(All the consumers), 
    \STATE buffer: PriorityQueue, writeSize: Integer
    \WHILE{Messages in Consumers}
        \FOR{each consumer in Consumers}
          \STATE message = consumer.GetMessage()
          \STATE buffer.put(message)
          \IF{buffer.size() $\geq$ writeSize}
            \STATE Deliver messages from the buffer till continuous messages are available.
          \ENDIF
        \ENDFOR
    \ENDWHILE
\end{algorithmic}
\end{algorithm}

The Aggregator and Sorter Algorithm \ref{Aggregator and Sorter} plays a crucial role in managing the flow of messages, optimizing the order of processing, and improving the overall system's performance.

\subsection{\textbf{Single Consumer within a Consumer Group Algorithm}}

One straightforward approach to preserve message order is to streamline message delivery. This can be accomplished either by creating a single partition for each topic or by assigning a single consumer within a consumer group to all partitions of a topic. However, opting for a single partition per topic lacks scalability because the broker handling leader partitions can become easily overwhelmed with increased network traffic. Consequently, we propose the adoption of a single consumer as a more viable alternative.

In the case of a single consumer, we employ a round-robin polling strategy across partitions, ensuring the delivery of messages in the order they arrive. The message key field plays a crucial role in determining message sequence. Out-of-order messages are temporarily stored within the consumer and subsequently delivered in the correct order upon receiving the missing sequentially numbered messages.

\begin{algorithm}
\label{Single Consumer within a consumer group}
\caption{Single Consumer within a consumer group}
\begin{algorithmic}

\STATE Initialize message order preservation strategy

\IF{Using a single partition per topic}
    \STATE Create a single partition for each topic
\ELSIF{Assigning a single consumer within a consumer group to all partitions}
    \STATE Assign a single consumer to handle all partitions of the topic
\ELSE
    \STATE Choose an alternative approach
\ENDIF

\IF{Opting for a single consumer}
    \STATE Initialize a round-robin polling strategy
    \FOR{Each message in partitions}
        \STATE Poll messages in a round-robin sequence
        \STATE Use the message key field to determine the message sequence
        \IF{Message is out-of-order}
            \STATE Buffer the out-of-order message
        \ENDIF
        \IF{Received missing sequentially numbered messages}
            \STATE Deliver out-of-order messages in the correct order
        \ENDIF
    \ENDFOR
\ELSE
    \STATE Choose an alternative approach
\ENDIF

\end{algorithmic}
\end{algorithm}

Single Consumer within a Consumer Group Algorithm \ref{Single Consumer within a consumer group} addresses the challenges associated with message handling in distributed systems.

\subsection{\textbf{Batch Commit and Broadcast Protocol Algorithm}}

This approach suggests preserving order by employing a consensus algorithm independently among producers and consumer groups. To accomplish this, we introduce a global batch size at the producer level for ordered messages. During a single poll operation, the consumer receives messages in multiples of this batch size.

\begin{algorithm}
\label{Batch Commit and Broadcast Protocol Algorithm}
\caption{Batch Commit and Broadcast Protocol Algorithm}
\begin{algorithmic}

\STATE \textbf{Producers' Role:} Producers employ Raft consensus algorithms to assign a batch number to a group of messages and then write them into the broker sequentially, following a round-robin approach. This batch number ensures a uniform sequence identifier across all system components.

\STATE \textbf{Partitioning Strategy:} Instead of employing key-based partition allocation, we will utilize the Round Robin Partitioner (available in Kafka) to distribute messages across partitions.

\STATE \textbf{Consumers' Role:} Consumers will adopt the atomic broadcast protocol to ensure the sequential delivery of messages, guided by batch numbers. While consumers can continue to poll multiple batches of messages, during delivery to the application, they will prioritize delivering them in accordance with the batch number sequence. Subsequently, they will broadcast the information about the next batch to be delivered or the last batch number that was dispatched. When a consumer possesses messages from the next batch, it will deliver them to the client and inform all other consumers accordingly.

\end{algorithmic}
\end{algorithm}

Batch Commit and Broadcast Protocol Algorithm \ref{Batch Commit and Broadcast Protocol Algorithm} gives highly efficient Kafka streams that can provide global ordering of the messages without any new layer or bottleneck at any layer. Figure 3 explains the Architecture Design using Batch Commit and Broadcast Protocol.

\begin{figure}[htbp]
\includegraphics[width=90mm,scale=1.0]{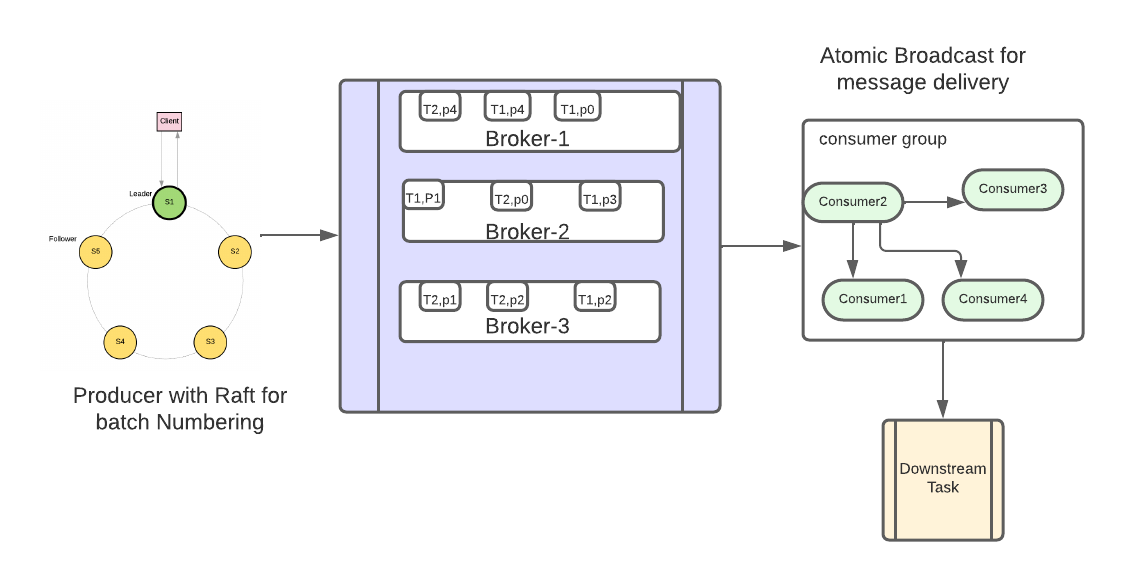}
\caption{Architecture Design using Batch Commit and Broadcast Protocol}
\end{figure}

\section{Experiments And Results}

In this research, we undertook a systematic experimental comparison of the framework, which was constructed following the designs delineated in earlier sections. We aimed to contrast the latency and throughput of our developed system with the inherent attributes of the native Kafka framework. Given that our approach functions as an overlay atop Kafka, it is anticipated that message delivery might exhibit augmented latency. Efforts were made to maintain consistent parameters across different design configurations wherever feasible.

We utilized a Macintosh system equipped with an 8-core CPU, segmented into 4 performance cores and 4 efficiency cores, complemented by an 8-core GPU and a 16-core Neural Engine for our experiments. To simulate a multi-producer and multi-consumer setup, we initiated several threads sharing a common group ID.

Within the context of the \textbf{Aggregator and Sorter design paradigm}, multi-threading was employed to simulate the simultaneous operations of multiple producers and consumers. When a client request is received, the producer engages a lock via a distributed lock service, subsequently generating a sequential token. This locking mechanism is critical, guaranteeing the uniqueness and orderly sequence of tokens, thereby preventing any duplication or misordering. 

Following this, the producers relay their respective messages to the broker, where these messages are stored with their keys designated by the sequence token ID. Upon retrieval from the broker, the consumer places the message in a distributed queue structured to uphold the message sequence and facilitate the delivery of organized messages. It is imperative to note that message delivery is initiated only after reaching a predefined buffer size, ensuring a globally sequenced batch dispatch.

\begin{figure}[htbp]
\includegraphics[width=90mm,scale=0.8]{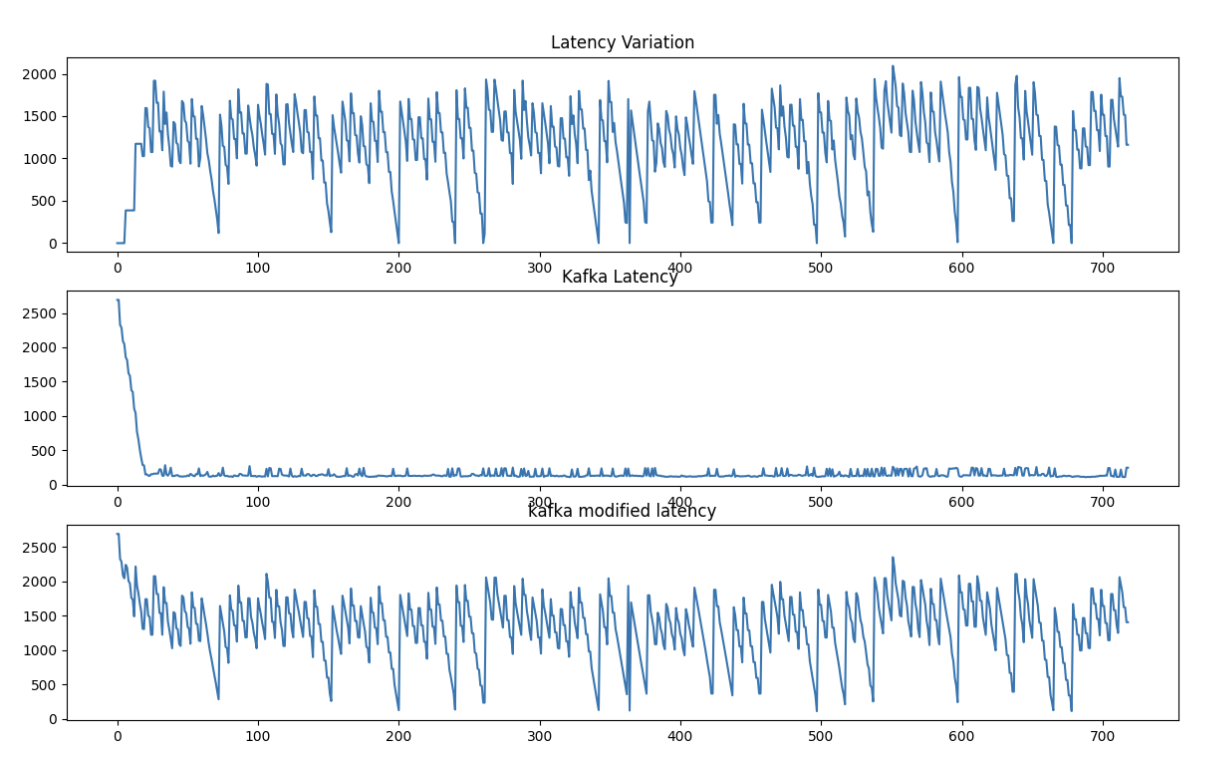}
\caption{Multi Consumer Aggregator and Sorter Design Performance}
\end{figure}

In order to assess the efficacy of the \textbf{Multi Consumer Aggregator and Sorter Design} implementation, we conducted an experimental study involving the transmission of a burst of 700 messages. Our analysis revealed that the average latency per request when employing the native Kafka system was approximately 6.9 milliseconds, whereas the utilization of the modified Kafka wrapper resulted in an average latency of 60 milliseconds.

Fig. 4 displays a graphical representation of the relationship between Request ID and their corresponding latency, measured in hundredths of a second $\frac{1}{100}^{th}$ of a second). Notably, as depicted in Fig. 4, an observation can be made regarding the latency disparity between the native Kafka system and the modified Kafka implementation, amounting to approximately 20 milliseconds.

For a \textbf{single consumer design paradigm}, a single thread was used to read from all the partitions. A local buffer is maintained that is responsible for sorting the messages received based on the message key and delivering them to the downstream process in a globally sorted order. 

\begin{figure}[htbp]
\includegraphics[width=90mm,scale=0.8]{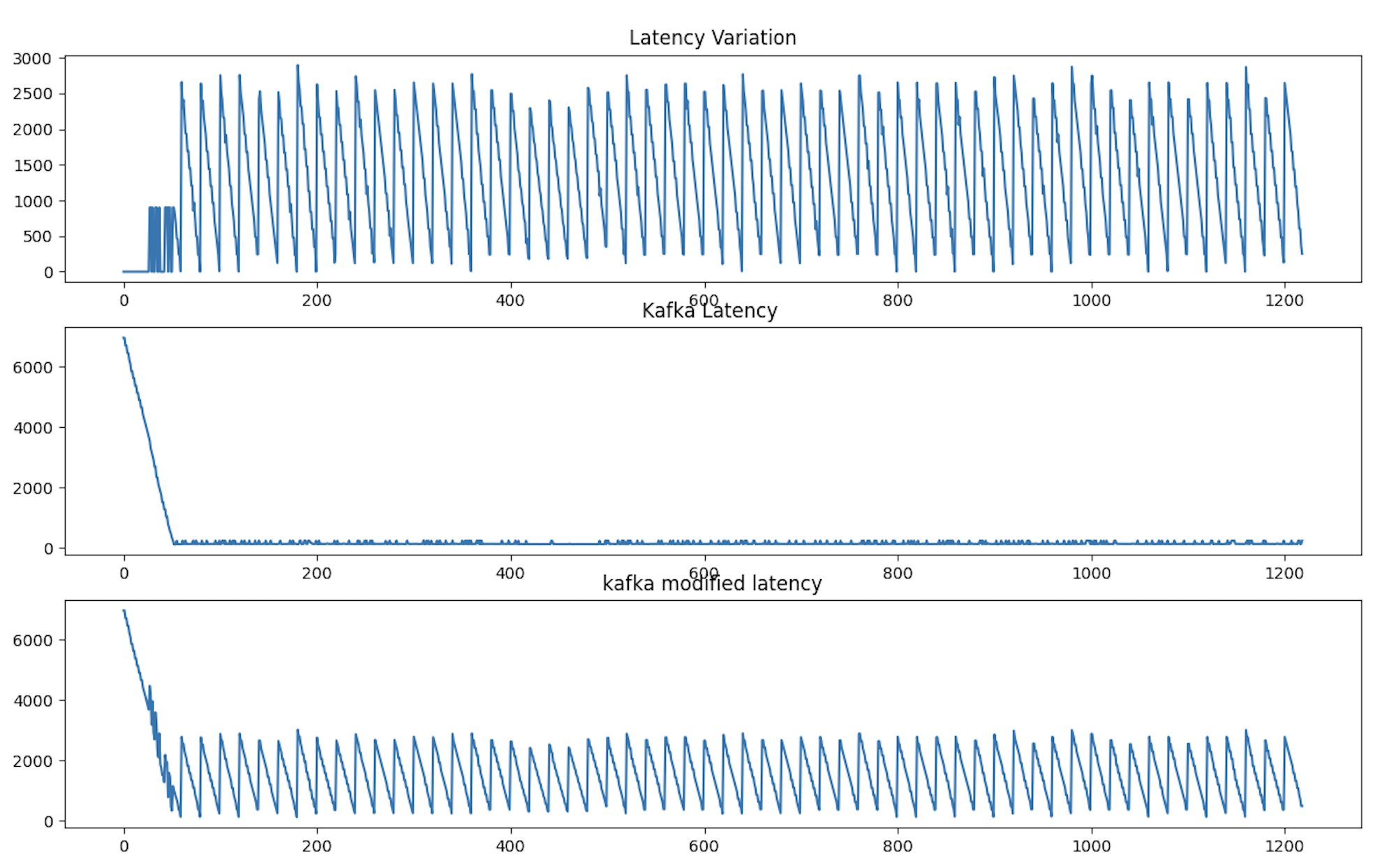}
\caption{Single Consumer Design Performance}
\end{figure}

The average latency per request for a \textbf{single consumer design} with 3 partitions was observed around 16ms. The difference in the latency between native Kafka and modified Kafka is around 9 ms. The average request latency for a \textbf{single consumer design} utilizing three partitions measured approximately 16ms. Notably, there is a discernible latency disparity of roughly 9ms between the unaltered Kafka setup and the customized Kafka setup.

For the implementation of the \textbf{batch commit and broadcast protocol}, we've incorporated the Raft algorithm on the producer side to generate sequential token IDs. Instead of assigning a token ID to each individual message, we allocate it to a batch, which can be configured to a specific size in the native Kafka environment. Correspondingly, the consumer reads from a broker with an identical batch size to that of the producer. To facilitate message delivery to downstream applications, we've employed an atomic broadcast protocol with a built-in timeout mechanism.

Upon reading messages from the broker, the consumer patiently awaits a broadcast message that conveys the sequence ID of the next batch to be committed. In the event of a timer expiration, the consumer broadcasts the lowest batch sequence number available in its buffer. Upon receiving this broadcast message, other consumers also respond with their lowest batch sequence numbers. Each consumer independently computes the lowest batch sequence number, which determines the order of commitment. The consumer bearing the lowest batch sequence number proceeds to deliver the message and initiates the broadcast of the subsequent batch sequence number scheduled for commitment.

\begin{figure}[htbp]
\includegraphics[width=90mm,scale=0.8]{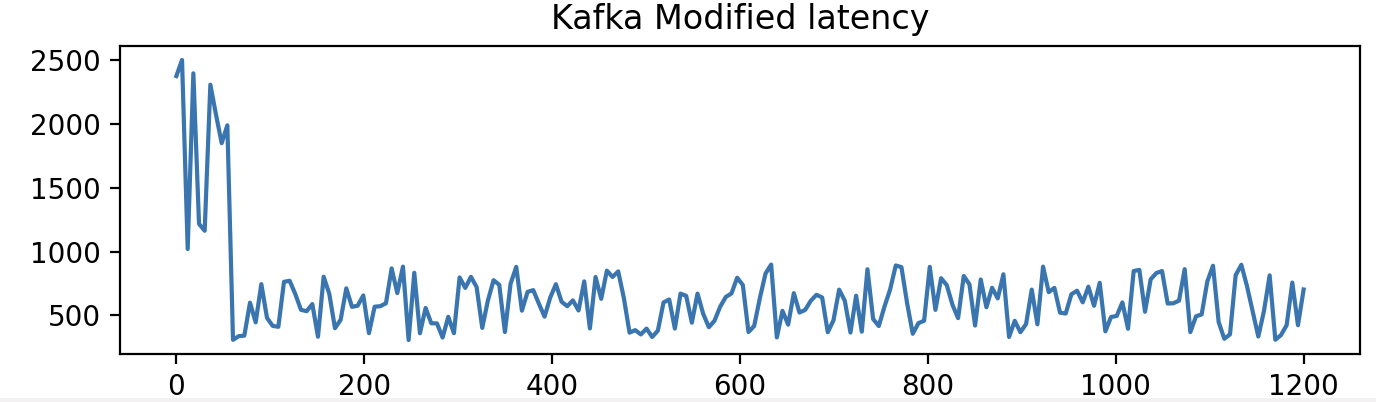}
\caption{Batch commit and Broadcast Protocol Performance}
\end{figure}

We conducted a thorough examination of the average latency per request within the \textbf{batch commit and broadcast protocol} design. Our observations revealed that the average latency per request consistently registered at approximately 9.0 ms. In the context of performance comparison between native Kafka and the modified Kafka version, a discernible difference of approximately 2 ms became apparent.

While the \textbf{batch commit and broadcast protocol} design within the modified Kafka version offers certain advantages, it does introduce an incremental latency of 2 ms when contrasted with native Kafka. This information holds significant relevance for system architects and developers who prioritize real-time data processing and are actively assessing the trade-offs between system performance and necessary modifications. Refer to Table 1 for Performance Analysis of all three designs.

\begin{table}[htbp]
\centering
\caption{Experimental Results}
\begin{tabular}{|p{3cm}|p{2cm}|p{2cm}|}
\hline

\textbf{Design Paradigm} & \textbf{Native Kafka Latency (ms)} & \textbf{Modified Kafka Latency (ms)} \\
\hline
Aggregator and Sorter & 6.9 & 60 \\
\hline
Single Consumer (3 partitions) & 16 & 25 \\
\hline
Batch Commit and Broadcast Protocol & 9 & 11 \\
\hline
\end{tabular}
\end{table}

\section{Conclusion And Future Work}

From our tests, it became evident that a single consumer outperforms the Aggregator and Sorter for the given message burst size. Hence, for applications with a lower frequency of messages and partitions, the single consumer emerges as the superior choice. 

However, as the message frequency and number of partitions rise, the performance of the Aggregator and Sorter improves. Notably, the batch commit and broadcast protocol demonstrated reduced latency in generating sequence IDs compared to distributed locks. The inclusion of a buffer in the initial two designs creates a consistent latency, as we have to wait for the sorter and aggregator layers to attain a specific buffer size. Interestingly, we noted enhanced performance using the atomic broadcast protocol with re-transmission, possibly due to real-time message delivery as opposed to buffering with a batch size greater than one. While the atomic broadcast protocol offers speed, it introduces the challenge of overseeing group membership, a task managed by Kafka's zookeeper.

We evaluated our current models within a multi-threaded environment. To conduct a more comprehensive performance analysis, we are contemplating replicating these designs within a distributed framework spanning diverse geographical locations. While our present testing centers around the latency per request for a single batch size, our future efforts are geared towards exploring additional metrics, including the throughput measured in requests processed per second, and investigating the impact of varying batch sizes.

Our designs function as wrappers around Kafka, but our aspiration is to transform them into libraries suitable for applications that require global message ordering. Furthermore, our designs possess the capability to prioritize messages, contingent on the availability of a token-generating algorithm that adheres to specific priority guidelines.

\section*{Acknowledgment}

The authors would like to express their gratitude to those who have provided feedback and support throughout the research process. Furthermore, we acknowledge that a significant portion of the content presented in this paper has been derived from our previously published arXiv preprint titled "Distributed Kafka Clusters: A Novel Approach to Global Message Ordering" \cite{kumar2023distributed}.
\nocite{*}
\bibliographystyle{IEEEtran}
\bibliography{IEEEexample}
\end{document}